\documentclass[modern]{aastex61}
\usepackage{amsmath}

\newcommand{\msun}{\mbox{$M_{\sun}$}\,}

\shorttitle{Magnetic field in RCW~106 mini-starburst}
\shortauthors{Tamaoki et al.}

\begin{document}

\title{Magnetic stability of massive star forming clumps in RCW~106}

\author{Shohei Tamaoki}
\affiliation{Graduate School of Natural Sciences, Nagoya City University, Mizuho-ku,
Nagoya 467-8501, Japan}
\author{Koji Sugitani}
\affiliation{Graduate School of Natural Sciences, Nagoya City University, Mizuho-ku,
Nagoya 467-8501, Japan}
\author{Quang Nguyen-Luong}
\affiliation{IBM Canada, 120 Bloor Street East, Toronto, ON, M4Y 1B7, Canada}
\author{Fumitaka Nakamura}
\affiliation{National Astronomical Observatory of Japan, Mitaka, Tokyo 181-8588, Japan}

\author{Takayoshi Kusune}
\affiliation{National Astronomical Observatory of Japan, Mitaka, Tokyo 181-8588, Japan}
\author{Takahiro Nagayama}
\affiliation{Kagoshima University, 1-21-35 Korimoto, Kagoshima 890-0065, Japan}
\author{Makoto Watanabe}
\affiliation{Department of Applied Physics, Okayama University of Science, 1-1 Ridai-cho, Kita-ku, Okayama 700-0005, Japan}
\author{Shogo Nishiyama}
\affiliation{Miyagi University of Education, 149 Aramaki-aza-Aoba, Aobaku, Sendai, Miyagi 980-0845, Japan}
\author{Motohide Tamura}
\affiliation{Department of Astronomy, The University of Tokyo, 7-3-1, Hongo, Bunkyo-ku, Tokyo, 113-0033, Japan}
\affiliation{Astrobiology Center of NINS, 2-21-1, Osawa, Mitaka, Tokyo 181-8588, Japan}
\affiliation{National Astronomical Observatory of Japan, Mitaka, Tokyo 181-8588, Japan}

\begin{abstract}
The RCW~106 molecular cloud complex is an active massive star-forming region where a ministarburst is taking place. 
We examined its magnetic structure by near-IR polarimetric observations with the imaging polarimeter SIRPOL on the IRSF 1.4 m telescope.  
The global magnetic field is nearly parallel to the direction of the Galactic plane and the cloud elongation. 
We derived the magnetic field strength of $\sim100$--$1600~\mu$G for 71 clumps with the Davis-Chandrasekhar-Fermi method. 
We also evaluated the magnetic stability of these clumps and found massive star-forming clumps tend to be magnetically unstable and gravitationally unstable. 
Therefore, we propose a new criterion to search for massive star-forming clumps.  
These details suggest that the process enhancing the clump density without an increase of the magnetic flux is essential for the formation of massive stars and the necessity for accreting mass along the magnetic field lines. 

\end{abstract}


\section{Introduction} \label{sec:intro}
\begin{figure*}[htbp]
\begin{center}
\includegraphics[width=15cm]{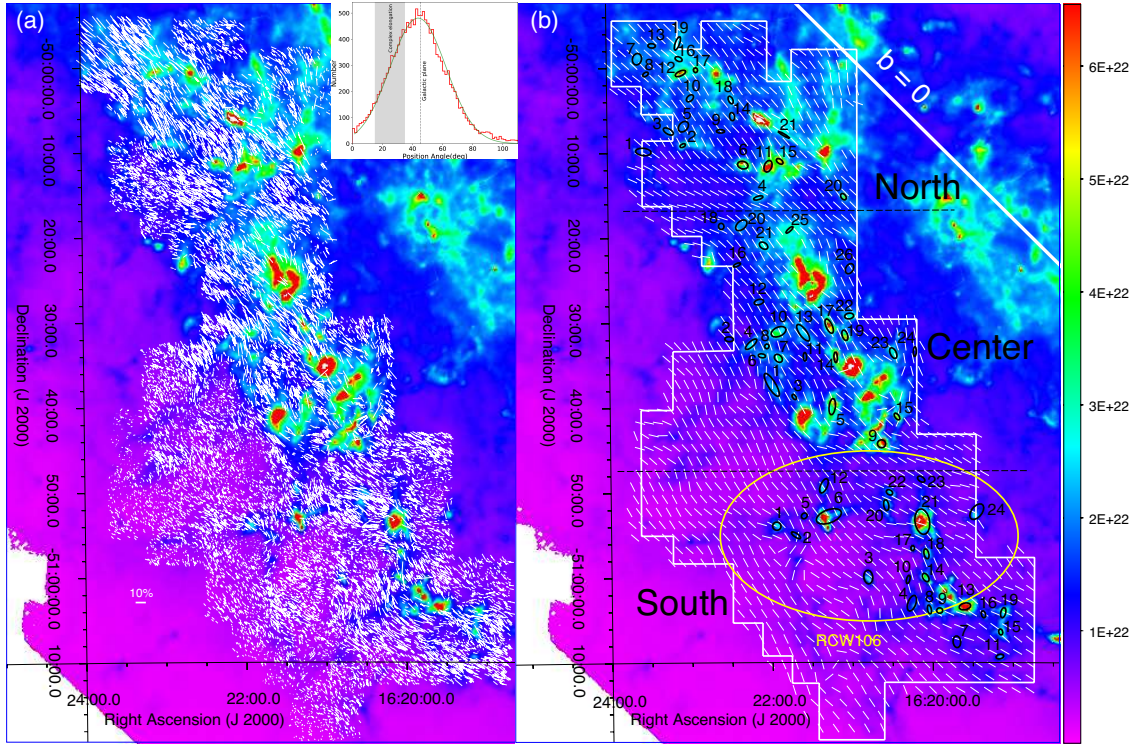}\\
\caption{{\bf (a):} $H$-band polarization vector map overlaid on the $N_{\rm H_2}$ map. 
The histogram of the $H$-band polarization angles is shown at upper right with a Gaussian fitting line (green line). 
{\bf (b): } The average angles of the $H$-band vectors within 3$\arcmin \times 3\arcmin$ square with a grid spacing of 1.5$\arcmin$ are shown on the $N_{\rm H_2}$ map. 
Black ellipses with identification numbers, a yellow ellipse, and a white polygon show $astrodendro$ clumps, the extent of the RCW~106 HII region, and the area of our near-IR polarimetry, respectively.  
\label{fig:f1}}
\end{center}
\end{figure*}

Magnetic fields are believed to play an essential role in star formation at all scales because the interstellar medium is generally magnetized \citep{crutcher12}. 
However, whether they play deconstructive or constructive roles in the evolution of stars and molecular clouds is not entirely understood. 
From the theoretical point of view, magnetic field is a process that can be distinguished in massive and low-mass star formation \citep[see][]{Shu_etal1987}. 
Magnetic fields in magnetically subcritical clumps support the clumps from collapsing gravitationally under conservation of magnetic flux and therefore only allow the cloud to form low-mass stars with ambipolar diffusion that advances steady and slow instead of massive stars that require drastic collapsing. 
Conversely, magnetically supercritical ones would generate the high-mass core needed for massive star formation thanks to the onset of relatively rapid contraction.

At a distance of 3.6~kpc \citep{lockman79}, the RCW~106 molecular cloud complex is a very active massive star-forming region, so active that it is classified as a ministarburst site. 
Its 83 pc long structure is located in the Scutum--Centaurus arm and is elongated approximately in the ENE--WSW direction ($\sim$25\degr ~from north to east). 
This $6\times10^6\,\msun$ complex is powered by the giant H$_{\rm II}$ region RCW~106 \citep{Nguyen_2015}, one of the largest and brightest HII regions in the Milky Way. 
The giant HII region RCW~106 hosts a cluster with a mass of $\sim 10^3\,\msun$\ and Lyman continuum photon emission of $10^{50}$ s$^{-1}$, likely originating from dozens of OB-type stars ($M>8~\msun$) \citep{lynga64} or radio continuum photon emission that is responsible for 54 OB stars \citep{Nguyen_2015}. 
High density tracers such as CS, HCO$^{+}$, HCN, HNC, NH$_{3}$ emission lines revealed a large sample of cold clumps and these clumps coincide with 1.2~mm dust clumps \citep{mooker04}, which are sites of massive star formation or are gravitationally unstable clumps and potentially forming stars  \citep{bains06,wong08,lo09,lowe14}. 

Although being a famous massive star-forming complex, its magnetic field structure remains unknown. 
Therefore, we observed the polarized starlight in near-infrared bands 
using the imaging polarimeter SIRPOL (polarimetry mode of the SIRIUS camera; \citealt{Kandori_2006}) mounted on the Infrared Survey Facility (IRSF) 1.4 m telescope at the South African Astronomical Observatory. 

\section{OBSERVATIONS} \label{sec:obs}
\subsection{Polarimetric observations with SIRPOL}

Magnetic fields can be revealed at near-IR by starlight polarization due to interstellar grain alignment based on radiative processes \citep[see][]{Andersson_2015}.
Near-IR imaging polarimetry toward RCW~106 was made on April and May 2017, January, July, and August 2018 with SIRPOL/SIRIUS on IRSF.
The camera has simultaneous observation capability at $JHK_{\rm{s}}$ bands using three 1024 $\times$ 1024 HgCdTe arrays, $JHK_{\rm{s}}$ filters, and dichroic mirrors (\citealt{Nagashima_1999}; \citealt{Nagayama_2003}). 
The field of view at each band is $\sim$7$\farcm$7 $\times$ 7$\farcm$7 with a pixel scale of 0$\farcs$45. 
We have observed 54 fields in total.  
For each field, we obtained ten dithered exposures, each of 15 seconds long, at four waveplate angles (0$\degr$, 22$\fdg$5, 45$\degr$, and 67$\fdg$5 in the instrumental coordinate system) and repeated it six times. 
Thus, the total exposure time was 900 seconds for each wave-plate angle. 
The seeing size ranged from $\sim$1\farcs5 to 2\farcs3 at $H$ band. 
Twilight flat-field images were obtained at the beginning and/or end of the observations. 
Standard image reduction procedures were applied with IRAF/PyRAF. 
Aperture photometry was executed at $J$, $H$, and $K_{\rm s}$, with an aperture radius of $\sim1$ FWHM corresponding to the seeing size.  
The 2MASS catalog \citep{Skrutskie_2006} was used for photometric/astrometric calibration. 
Only the sources with photometric measurement errors of less than 0.1 mag were used for analysis.
The Stokes parameters were calculated as  
$ I = (I_0+I_{22.5}+I_{45}+I_{67.5})/2$ and $q = (I_0 - I_{45})/I, u = (I_{22.5} - I_{67.5})/I$, where $I_0$, $I_{22.5}$, $I_{45}$, and $I_{67.5}$ are the intensities at four wave plate angles. 
The Stokes parameters were converted into the equatorial coordinate system $(q',u')$ with a rotation of 105$\degr$ \citep{Kandori_2006, Kusune_2015}. 
The degree of polarization $P$ and the polarization angle $\theta$ were calculated as $\theta$ = (1/2)atan$(u'/q')$ and $P = (q^2 + u^2 )^{1/2}$. 
The errors in polarization ($\Delta P$ and $\Delta \theta$) were derived from the photometric errors. 
We adopted the measurable polarization limit of $\sim0.3\%$ \citep{Kandori_2006} and $\Delta P = 0.3\%$ was assigned to the sources of $\Delta P < 0.3\%$. 
The degrees of polarization were debiased as $P_{\rm{debias}} =(P^2 - \Delta P^2)^{1/2}$ \citep{Wardle_1974}. 
Because of the high polarization efficiencies of 95.5\% at $J$, 96.3\% at $H$, and 98.5\% at $K_{\rm s}$ \citep{Kandori_2006}, no particular corrections were applied further.

\subsection{Archival Data} \label{sec:arc}
The $Herschel$ Science Archival SPIRE/PACS data were used to obtain the H$_2$ column density map.  
First, we convolved the 350/250/160 $\mu$m images to the 500 $\mu$m image resolution, 36$\arcsec$. 
Then, we derived the spectral energy distribution (SED) at each pixel by SED fitting using the four images described above, in the same way as \citet{Konyves_2010}.
We adopted the dust opacity per unit mass, 
$\kappa_{\nu} = 0.1(\nu/1000\rm{GHz})^{\beta}$ \rm{cm}$^2$/\rm{g}, 
where $\beta = 2.0$. 

We fitted only pixels where signals are detected more than 3 rms in all four bands. 
The rms was measured around the reference area (RA$_{\rm{J2000}}$ = 16:19:19.64, DEC$_{\rm{J2000}}$=-51:45:36.1). 
We obtained a column density ($N_{\rm H_2}$) map with a higher resolution of 18$\arcsec$ using an equation of 
$N_{\rm H_2} = I_\nu/[B_\nu(T_{\rm d})\kappa_{\nu}\mu m_{\rm H}]$, where $I_\nu$ is the 250 $\mu$m non-convolved brightness, 
$B_\nu (T_{\rm d})$ is the Planck function at the dust temperature $T_{\rm d}$ derived by the SED fitting (mean: $22.5\pm1.9$ K, range: $\sim15$--$42$ K), $\mu$ is the mean molecular weight of 2.8, and $m_{\rm H}$ is the hydrogen atom mass. 
The obtained $N_{\rm H_2}$, when convolved to the 36$\arcsec$ resolution, is consistent with that of the SED fitting within 10\%.

We also use the $^{13}$CO and C$^{18}$O ($1$--$0$) cube from the Three-mm Ultimate Mopra Milky Way Survey (ThrUMMS) survey \citep{Barnes_2015}.

\section{ANALYSIS} \label{sec:ana}
\subsection{Clumps in RCW~106 cloud complex}
\subsubsection{Clump identification}
Figure~\ref{fig:f1} shows prominent large-scale structures that comprise of clumpy sub-structures. 
The cloud complex has a global elongation angle in PA $\sim25.0\degr$, slightly different from the Galactic plane angle of $\sim 45\degr$. 

We identify the clumpy structures with the $astrodendro$ package\footnote{\url{https://dendrograms.readthedocs.io/en/stable/index.html}} \citep{Rosolowsky_2008}.
The $astrodendro$ is an unsupervised hierarchical clustering method that build up cluster in a tree-like structure where each node represents a leaf, structure that has no sub-structure, or a branch, structure that has successor structure. 
It also computes physical properties of detected leaves that we regard as clumps (e.g., $S$: leaf area, $\sum_{S} N_{\rm H_2}$: sum of $N_{\rm H_2}$ over $S$).

Because the background column density gradually increases from South to North, i.e, approaching the Galactic plane, we separated the cloud complex into three regions: North, Center, and South. 
We estimated the background level $N_{\rm min}$ (minimum value to be considered) in each region as the flat and lowest column density level, which are $8.0\times 10^{21}$cm$^{-2}$, $6.0\times 10^{21}$cm$^{-2}$ and $4.0\times 10^{21}$cm$^{-2}$ for North, Center, and South, respectively. 
We set 60 pixels, corresponding to 1.5 $\times$ 18$\arcsec$, as the minimum number of pixels needed to define a structure as a leaf. 
We set $2\times 10^{21}$cm$^{-2}$ as the minimum delta parameter (minimum height to be defined as a leaf), which is roughly five times of the dispersion of $N_{\rm H_2}$ measured in the deemed background areas, not to detect too small structures ($\lesssim30\msun$). 
We calculated the mean column densities ($\overline{N}_{H_2}=\sum_S N_{\rm H_2}/S$) for each leaf. 
For analysis, we included only the clumps with $\overline{N}_{H_2} - N_{\rm min} > 7.0\times 10^{21}$cm$^{-2}$, which is the threshold in the regions such as cores can form \citep{Konyves_2015}, and assigned their mean net column densities as $\overline{N}_{\rm net} = \overline{N}_{H_2} - N_{\rm min}$.  
We calculated its sphere-equivalent radius as $R = \sqrt{S/\pi}$.
Then, the mass and mean volume density are evaluated as $M = \mu m_{\rm H}\overline{N}_{\rm net}S$ and $\rho = (3/4R)\mu m_{\rm H}\overline{N}_{\rm net}$, respectively.
The $R$ range is $\sim$0.46--2.29 pc and the mean $R$ is $0.77\pm0.32$ pc.
The $M$ range is $\sim$230--10600 $\msun$(Table~\ref{tab:clumpprop}).

\subsubsection{Star-forming properties of clumps}
We search for signs of star formation in clumps via the existence of mid-IR emission using the AllWISE source catalog \citep{Wright_2010} and the $Spitzer$ 24 $\mu$m images for clumps without AllWISE sources. 
For clumps without any 24 $\mu$m point source, we adopted three times the sum of standard deviation within $13\arcsec$ from the clump center as the detection limit. 
We classified the clumps into three groups: mid-IR bright clumps that have AllWISE sources, mid-IR faint clumps that are detected only in $Spitzer$ emission, mid-IR quiet clumps that have neither detection.

To understand the star formation activities of the clumps, we estimate their bolometric luminosities $L_{\rm bol}$ (Table~\ref{tab:clumpprop}),  because they are direct scales of the star formation rates \citep{inoue00}.
$L_{\rm bol}$ of AllWISE sources associated with the clumps are estimated based on the 12 and 22 $\mu$m flux from the AllWISE catalog and on the 70 and 160 $\mu$m flux from the PACS Point Source Catalog, following \citet{Chen_1995}. 
$L_{\rm bol}$ of bright sources that are saturated on the $Spitzer$ 24 $\mu$m or $WISE$ 22 $\mu$m images are estimated based on the IRAS point source flux using the method of \citet{Carpenter_2000}.  
For sources undetected in 70~$\mu$m or 160~$\mu$m, we use $F_{\rm mid-IR}$ as the proxy because we found a linear relation ${\rm log}_{\rm 10}(L_{\rm bol}/L_{\rm \sun})=(0.60\pm0.08)\times {\rm log}_{\rm 10}\left(F_{\rm mid-IR}/{\rm Jy}\right) + (3.3 \pm 0.1)$ with a linear fitting for the sources whose $L_{\rm bol}$ were estimated.
The source with a luminosity of $\ga 10^{3.8}$ $L_{\sun}$ is considered to be a massive B2 type star or more massive \citep{stahler05}. 

\subsection{Magnetic field in clumps}
\subsubsection{Polarization vectors}

\begin{figure}[ht!]
\begin{center}
\plotone{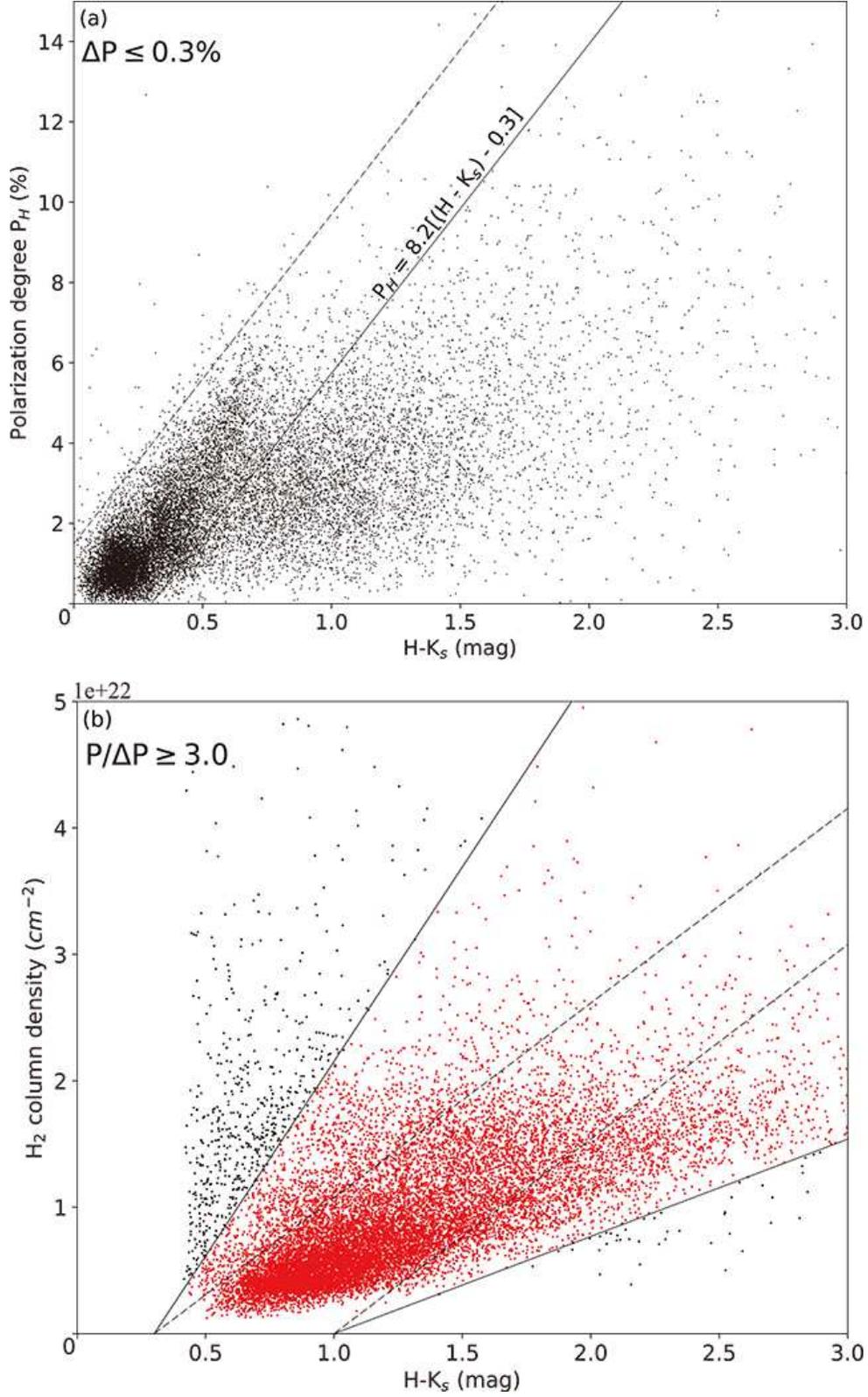}
\caption{(a) Polarization degree versus $H-K{\rm s}$ magnitude diagram for good polarimetric sources with $\Delta P \leq 0.3\%$. 
The solid line, which has the same slope as the dashed line, is our determined upper limit.  
See the details in Section 3.2.1. Here, we adopted $E (H - K_{\rm s}) = 0.065A_{\rm v}$ \citep{Cohen_1981}. 
(b) H$_2$ column density versus $H-K_{\rm s}$ color for the sources that are below our upper limit and have $P$ of $P/\Delta P \geq 3$. 
The dashed lines are the H$_2$ column densities expected from the color excess. 
See more details in Section 3.2.1.
\label{fig:f2}}
\end{center}
\end{figure}
In this Letter, we derive only the polarization vector map in $H$-band because it detected most sources associated with RCW 106 among the three bands. 

First, on the $P_{\rm H}$ vs ($H-K_{\rm s}$) diagram (Figure~\ref{fig:f2}a), we determined the upper limit to remove outlier sources with too high polarizations, such as stars with intrinsic polarization or polarization due to scattering off nebulosity \citep[e.g.,][]{Jones_1989}, using only the sources with good polarization accuracy ($\Delta P \leq 0.3\%$). 
We approximately estimated a upper threshold line to separate the outliers from the good measurements (a dashed line with a slope of 8.2).
The expected background source colors ($H-K_{\rm s}$)$_0$ without the reddening by the RCW~106 cloud have a range of ($H-K_{\rm s}$)$_0$ $\sim$ 0.3--1.0 (the model of Galactic IR point sources; \citealt{Wainscoat_1992}). We defined the upper limit as the linear line with a slope of 8.2 and passing through the point of ($H-K_{\rm s}$ = 0.3, $P$=0). 
Thus, we selected only the sources with polarization degree $P_H\leq8.2[(H-K_{\rm s}) - 0.3]$ and 
$P/\Delta P \geq 3$ (i.e., $\Delta\theta \lesssim 10\degr$) for analysis. 
The sources of the concentration at the lower left part of Figure~\ref{fig:f2}a appear to be foreground sources and mostly have optical counterparts in the DSS2 red image.
To remove the influence of foreground polarization, we estimated the mean $q'$ and $u'$ of these foreground sources with $d$=3.0--3.5 kpc \citep[referred from Gaia DR2;][]{gaia} and subtracted them from the $q'$ and $u'$ of the selected sources.

Second, considering ($H-K_{\rm s}$)$_0$ = 0.3--1.0 and the errors of $E(H-K_{\rm s})=[(H - K_{\rm s}) - (H - K_{\rm s})_0]$, we include only the sources that have their $H-K_{\rm s}$ excesses consistent with the $N_{\rm H_2}$ (red dots on Figure~\ref{fig:f2}b), within a factor of 2 that is the $N_{\rm H_2}$ uncertainty, which originates from $\kappa_{\nu}$ \citep{Konyves_2010}. 
The sources with too high $N_{\rm H_2}$ against their $E(H-K_{\rm s})$ would be foreground sources, while those with too low $N_{\rm H_2}$ would not sample the magnetic field of RCW~106. 
To calculate the column densities expected by the $E(H-K_{\rm s})$, we use the equations $E (H - K_{\rm s}) = 0.065A_{\rm v}$ \citep{Cohen_1981} and $N_{\rm E} = 1.0 \times 10^{21}A_{\rm v}$ \citep{Lacy_2017}, i.e, $N_{\rm E} =1.54 \times 10^{22}[(H - K_{\rm s}) - (H - K_{\rm s})_0]$ in units of ${\rm cm^{-2}}$. 

We present the $H$-band polarization vectors of the sources that meet the above criteria in Figure \ref{fig:f1}. 
The polarization vectors indicate that the global magnetic field direction seems to be nearly parallel to the Galactic plane and the global cloud elongation. 
The vector angle distribution was determined to be $43.7\pm18.7\degr$ with a Gaussian fit, of which the peak well agrees with the position angle of the Galactic plane.

\subsubsection{Magnetic strength of Clumps}

\begin{figure}[htbp!] 
\begin{center}
\plotone{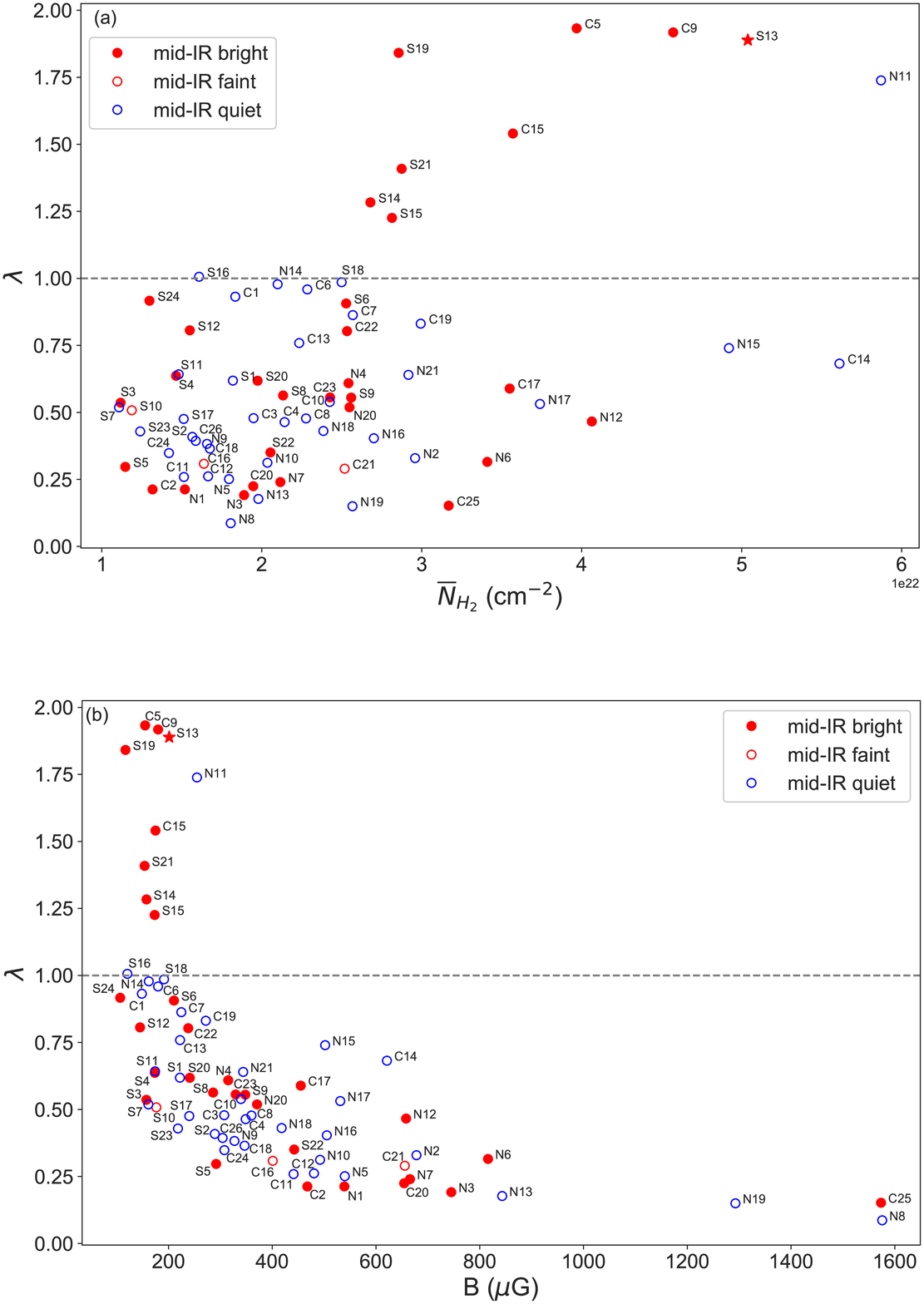}
\caption{Normalized mass-to-flux ratio $\lambda$ versus (a) mean column density, and (b) magnetic field strengths. }
\label{fig:f3}
\end{center}
\end{figure}

We derived the plane of the sky (POS) magnetic-field strength $B$ of each clump using the Davis-Chandrasekhar-Fermi method (\citealt{Davis_1951}; \citealt{Chandra_Fermi_1953}) modified by \citet{Ostriker_2001};

\begin{equation}
B =Q\sqrt{4\pi \rho} \frac{\sigma_v}{\sigma_{\theta}},
\end{equation}
where $\rho$ is the mean volume density of the clump, $\sigma_v$ is the mean velocity dispersion, $\sigma_{\theta}$ is the angular dispersion of the polarization vectors, and $Q$ is a correction factor of  0.5 ($\sigma_{\theta} <  25\degr$), introduced by \citet{Ostriker_2001} with their MHD simulations. 

We applied a single Gaussian fit to the $^{13}$CO cube data in the range of $V_{\rm LSR} =$ -80 to -30\,km s$^{-1}$ to determine the velocity dispersion $\sigma_{v}$ at each position of the $^{13}$CO cube data and obtained the mean  $\sigma_v$ within each clump.
For some clumps that have double peaks, double Gaussian fits were applied to the integrated $^{13}$CO spectra. 
Because $^{13}$CO might sample not only clump but also inter-clump materials, we correct $\sigma_{v}$ by dividing by the mean $\sigma_{v}(^{13}{\rm CO})/\sigma_{v}(\rm C^{18}O)$. 
We derived the mean $\sigma_{v}(^{13}{\rm CO})/\sigma_{v}(\rm C^{18}O)$ by taking only pixels that are detected in both lines ($T_{\rm peak} > 1.0$~K). 
Consequently, we obtained $<\sigma_{v}(^{13}{\rm CO})/\sigma_{v}{\rm (C^{18}O)}>=1.76\pm0.55$. 

To derive the angular dispersion $\sigma_{\theta}$ of the $H$-band polarization vectors, we adopted the method of \cite{Hildebrand_2009}. 
The angular difference is given as $\Delta \theta (l) \tbond \theta (x) - \theta (x+l)$, between the $N(l)$ pairs of vectors separated by the displacement $l$. 
The square of the angular dispersion function \citep[ADF; see also][]{Kobulnicky_1994} is expressed as follows: 
\begin{equation}
<\Delta \theta^2 (l)>\tbond {\frac{1}{N(l)}\sum^{N(l)}_{i=1}}[\theta (x) - \theta (x+l)]^2,
\end{equation}
and can be approximated as follows: 
\begin{equation}
<\Delta \theta^2 (l)>_{\rm tot}\simeq b^2 + m^2 l^2 + \sigma_{\rm M}^2 (l),
\label{eq:deltaPhi}
\end{equation}
where $b$, $m$, and $\sigma_{\rm M} (l)$ present the contributions of the turbulent dispersion, large-scale structure, and measurement uncertainties, respectively. 

We constructed the plot of squared ADF and $l$ and fit Equation\,\ref{eq:deltaPhi} to derive $b$ and $m$ \citep{Chapman_2011}, toward the clumps and its immediate surroundings. 
Following \cite{Hildebrand_2009}, we calculated $\sigma_{\theta}$ approximately as the ratio of the turbulent to large-scale magnetic field strength;  
\begin{equation}
\sigma_{\theta} = \frac{<B_{\rm t}^2>^{1/2}}{B_0} = \frac{b}{\sqrt{2-b^2}},
\end{equation}
where $B_0$ is a large-scale magnetic field, and $B_{\rm t}$ is a turbulent component. 
See the details in Section 3 of \citet{Hildebrand_2009}. 

We used the $H$-band vectors within $2\arcmin$, $\sim$ 2--3 times the clump $R$, from the clump center to make the fit.  
To avoid bad fitting, the clumps with the number of vectors $< 30$ were excluded. 
We exclude the clumps that have bad fits even if they have more than 30 vectors.
Seventy-one clumps are left for further analysis. 
We note that several clumps in the very high density areas are not included because the number of vectors of the background sources dose not satisfy our selection criterion. 
Finally, we obtain the magnetic field strengths of $\sim100$--$1600~\mu$G for 71 clumps. 

While \citet{Jones_2015} found that grain alignment becomes problem at $A_{\rm v}\ga 20$ in starless cores, \citet{Whittet_2008} suggested alignment enhancement around the embedded stars.  
Since the mid-IR bright clumps have embedded stars, such enhancement might have occurred and our analysis would be valid. 
Note that mid-IR quiet clumps have smaller R compared to the bright ones and there is a possibility that we do not properly estimate their magnetic fields, but their exterior's.

\begin{figure}[htbp!]
\begin{center}
\plotone{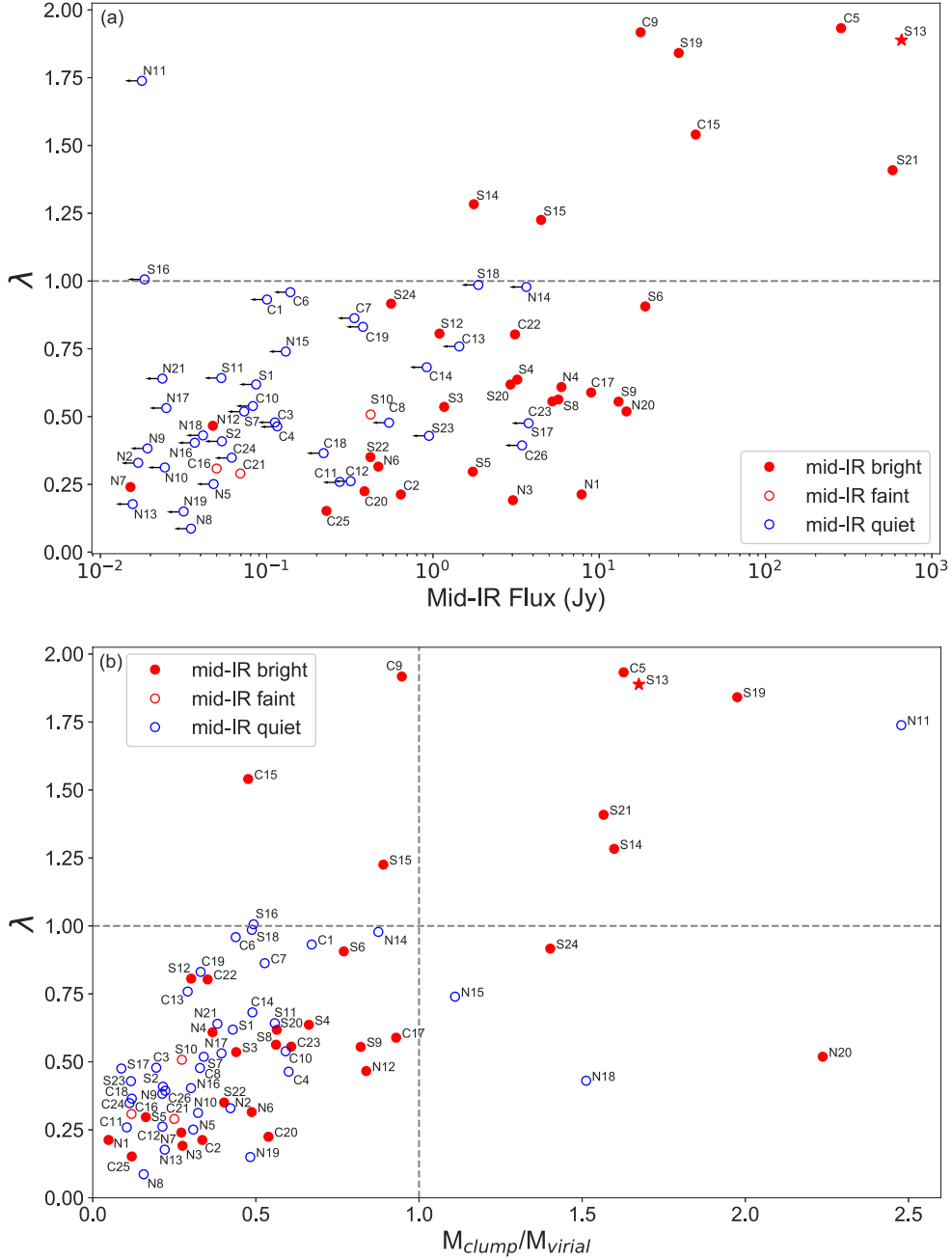}
\caption{Normalized mass-to-flux ratio versus (a) Mid-IR flux, and (b) $M_{\rm clump}/M_{\rm virial}$. 
For mid-IR quiet clumps,  the upper limits are indicated by small arrows (see the text). 
\label{fig:f4}}
\end{center}
\end{figure}

\section{DISCUSSIONS AND CONCLUSION} 
\label{sec:dis}
\subsection{Magnetic stability of clumps in RCW~106}
Magnetic field strengths derived from our measurements of the clumps in RCW~106 are about $100$--$1600~\mu$G and the distribution of the magnetic field strength $B$ is not much different among the different clump classes (Table \ref{tab:clumpprop} and Figure~\ref{fig:f3}). 
As mentioned Section 1, \citet{Shu_etal1987} predicted that the process of massive star formation is different from low-mass star formation. 
Magnetic fields in magnetically subcritical clumps prevent the clumps from collapsing gravitationally under conservation of magnetic flux. 
Magnetically supercritical clumps would generate the high-mass core needed for massive star formation because massive stars might require drastic collapsing. 

For a clump, the magnetic stability is quantifiable as the mass-to-magnetic-flux ratio $\lambda _{\rm obs}$ as
\begin{equation}
\lambda_{\rm obs} = \frac{\mu m_{\rm H} \overline{N}_{\rm net}}{B}
\end{equation}

or the normalized mass-to-magnetic flux ratio as
\begin{equation}
\lambda = \frac{\lambda_{\rm obs}}{\lambda_{\rm crit}},
\end{equation}
where $\lambda_{\rm crit} = 1/\sqrt{4\pi^2 G}$ is the stability criterion \citep{Nakano_Nakamura_1978}.
The clump is magnetically stable if $\lambda$ is equal to or less than 1, otherwise unstable. 

Sixty-two of the clumps are close to the critical condition or under the subcritical condition ($\lambda$ $\lesssim 1$) and $\lambda$ increases linearly with $\overline{N}_{\rm net}$, but inversely correlate with $B$ as expected (Figure~\ref{fig:f3}).
Almost all (36/37) mid-IR quiet clumps have $\lambda$ $\lesssim 1$ and only the clump N11 has more larger $\lambda$ of $> 1$. 
More than half (23/31) of the active star-forming (mid-IR bright) clumps are close to magnetically critical or subcritical, while 8 clumps are supercritical ($\lambda > 1$). 

We examined the correlation between $\lambda$ with $F_{\rm mid-IR}$ and ${M_{\rm clump}/M_{\rm virial}}$ in order to examine the relation between magnetic field instability with star formation activities and gravitational instability. 
$\lambda$ correlates almost linearly with log$_{10}(F_{\rm mid-IR}/\rm{Jy})$ and ${M_{\rm clump}/M_{\rm virial}}$ (Figure \ref{fig:f4}). 
The interesting feature is that the mid-IR {\it brighter} clumps tend to be magnetically supercritical (Figure \ref{fig:f4}a), especially the clumps with luminosity $\ga 10^{3.8}$ $L_{\sun}$ (flux $\ga 10$ Jy), which are classified as the massive star-forming clumps ($\ga$ 800 $M_{\odot}$). 
These facts suggest that massive stars tend to be formed in magnetically supercritical clumps. 
Figure \ref{fig:f4}b shows that mid-IR bright, or massive star-forming clumps are mostly magnetically and gravitationally unstable.

\subsubsection{Implication of magnetic fields on massive star formation in RCW~106}
Our results strongly suggest that massive star formation prefers to occur inside magnetically and gravitationally unstable clumps. 
The latter point is consistent with previous studies, both in observations \citep{nguyenluong16} and simulation \citep{howard16}. 
They claimed that massive star formation occurs in gravitationally unstable cloud complex rather than stable one. 
We therefore propose a new criteria for identifying massive star-forming clumps, which is:

\begin{equation*}
    \begin{array}{l}
     \text{Massive star}\\
     \text{forming clumps}\\
    \end{array}
    \iff 
  \left\{
    \begin{array}{l}
     \frac{M_{\rm clump}}{M_{\rm virial}}  > 1, \text{gravitationally unstable}\\
     \frac{\lambda_{\rm obs}}{\lambda_{\rm crit}} > 1, \text{magnetically unstable.}\\
    \end{array}
  \right.
\end{equation*}

Massive star-forming clumps are therefore lying in the upper-right of Figure~\ref{fig:f4}. 
Our results imply that massive star formation could more quickly occur in the magnetically unstable clumps. 
These suggest the importance of the process that enhances the clump density while not increases the magnetic flux for massive star formation, e.g., the buildup of molecular gas along the magnetic field. 
Naturally, supercritical clumps will arise in the agglomerated environments of clumps in large cloud complexes (\citealt{Shu_1987}; \citealt{Shu_etal1987}). 

\vspace{1cm}
This work was partly supported by JSPS KAKENHI Grant Numbers JP16H05730, JP17H01118. 
We thank Y. Nakajima for assistance in the data reduction with the SIRPOL pipeline package. S.T. thanks the Daiko Foundation for financial support of our research. 
M.T. is supported by JSPS KAKENHI grant Numbers JP18H05442, JP15H02063, JP22000005. 

\clearpage

\begin{deluxetable}{lccccrccrrrrcc}
\renewcommand{\arraystretch}{0.8}
\tablecaption{Properties of 71 clumps considered for analysis\label{tab:clumpprop}}
\tablewidth{0pt}
\tabletypesize{\tiny}
\tablehead{&&&&&&&&&&&&&\\\colhead{ID} & \colhead{RA (2000)} & \colhead{Dec (2000)} & \colhead{$\overline{N}_{\rm net}$} & \colhead{$R$} & \colhead{$M_{\rm cl}$} & \colhead{$\sigma_v(^{13}{\rm CO})$} & \colhead{$\sigma_v$(C$^{18}$O)\tablenotemark{a}} & \colhead{$\sigma_{\theta}$} & \colhead{$B$} & \colhead{$F_{\rm mid-IR}$} & \colhead{log($L$/$L_{\sun}$)} &\colhead{$\lambda$} & \colhead{mid-IR} \\
\colhead{}&\colhead{($\degr$)}&\colhead{($\degr$)}& \colhead{(10$^{22}$cm$^{-2}$)} & \colhead{(pc)} & \colhead{(\msun)} & \colhead{(km~s$^{-1}$)} & \colhead{(km~s$^{-1}$)} & \colhead{($\degr$)} & \colhead{($\mu$G)} & \colhead{(Jy)} & \colhead{} & \colhead{} & \colhead{source}}
\startdata
N1 & 245.890 & -50.164 & 1.52 & 1.00 & 1061 & 7.66 & 4.28 & 10.75 & 538 & 7.826 & 3.84\tablenotemark{b} & 0.38 & bright \\
N2 & 245.769 & -50.152 & 2.96 & 0.46 & 440 & 2.46 & 1.37 & 5.63 & 677 & $<$0.017 & $<$2.24\tablenotemark{d} & 0.59 & ... \\
N3 & 245.813 & -50.124 & 1.89 & 0.81 & 862 & 3.22 & 1.80 & 4.05 & 745 & 3.025 & 3.59\tablenotemark{b} & 0.34 & bright \\
N4 & 245.539 & -50.256 & 2.54 & 0.50 & 453 & 2.55 & 1.43 & 11.16 & 315 & 5.921 & 3.76\tablenotemark{b} & 1.09 & bright \\
N5 & 245.767 & -50.115 & 1.80 & 0.87 & 960 & 3.09 & 1.72 & 5.02 & 539 & $<$0.048 & $<$2.51\tablenotemark{d} & 0.45 & ... \\
N6 & 245.586 & -50.191 & 3.41 & 1.01 & 2431 & 3.63 & 2.03 & 5.02 & 815 & 0.469 & 3.10\tablenotemark{b} & 0.56 & bright \\
N7 & 245.906 & -49.982 & 2.12 & 0.94 & 1307 & 3.70 & 2.07 & 5.12 & 665 & 0.015 & 2.21\tablenotemark{b} & 0.43 & bright \\
N8 & 245.881 & -50.012 & 1.81 & 0.50 & 313 & 3.28 & 1.83 & 2.43 & 1575 & $<$0.035 & $<$2.43\tablenotemark{d} & 0.15 & ... \\
N9 & 245.653 & -50.125 & 1.66 & 0.50 & 287 & 2.69 & 1.50 & 9.20 & 327 & $<$0.019 & $<$2.27\tablenotemark{d} & 0.68 & ... \\
N10 & 245.746 & -50.060 & 2.04 & 0.55 & 432 & 2.55 & 1.42 & 6.10 & 492 & $<$0.024 & $<$2.33\tablenotemark{d} & 0.56 & ... \\
N11 & 245.509 & -50.194 & 5.87 & 0.85 & 2953 & 1.94 & 1.08 & 12.27 & 254 & $<$0.018 & $<$2.25\tablenotemark{d} & 3.11 & ... \\
N12 & 245.774 & -50.011 & 4.06 & 0.80 & 1825 & 2.69 & 1.50 & 5.65 & 657 & 0.047 & 2.51\tablenotemark{b} & 0.83 & bright \\
N13 & 245.861 & -49.956 & 1.98 & 0.60 & 492 & 3.16 & 1.77 & 4.19 & 843 & $<$0.016 & $<$2.22\tablenotemark{d} & 0.32 & ... \\
N14 & 245.618 & -50.096 & 2.10 & 0.54 & 420 & 1.55 & 0.86 & 11.61 & 161 & $<$3.644 & $<$3.64\tablenotemark{d} & 1.75 & ... \\
N15 & 245.473 & -50.184 & 4.92 & 0.59 & 1201 & 2.21 & 1.24 & 7.80 & 501 & $<$0.130 & $<$2.77\tablenotemark{d} & 1.32 & ... \\
N16 & 245.779 & -49.983 & 2.70 & 0.52 & 501 & 2.94 & 1.64 & 8.17 & 505 & $<$0.037 & $<$2.44\tablenotemark{d} & 0.72 & ... \\
N17 & 245.728 & -50.005 & 3.74 & 0.48 & 611 & 2.93 & 1.63 & 9.40 & 531 & $<$0.025 & $<$2.34\tablenotemark{d} & 0.95 & ... \\
N18 & 245.622 & -50.063 & 2.39 & 0.58 & 570 & 1.31 & 0.73 & 3.89 & 418 & $<$0.041 & $<$2.47\tablenotemark{d} & 0.77 & ... \\
N19 & 245.783 & -49.952 & 2.57 & 0.68 & 821 & 2.59 & 1.45 & 2.40 & 1292 & $<$0.032 & $<$2.40\tablenotemark{d} & 0.27 & ... \\
N20 & 245.278 & -50.254 & 2.55 & 0.66 & 765 & 1.18 & 0.66 & 3.85 & 370 & 14.573 & 4.00\tablenotemark{b} & 0.93 & bright \\
N21 & 245.460 & -50.132 & 2.92 & 0.50 & 505 & 2.66 & 1.49 & 11.50 & 343 & $<$0.024 & $<$2.32\tablenotemark{d} & 1.15 & ... \\
\hline
C1 & 245.500 & -50.622 & 1.83 & 1.37 & 2405 & 2.64 & 1.48 & 12.63 & 148 & $<$0.100 & $<$2.70\tablenotemark{d} & 1.67 & ... \\
C2 & 245.631 & -50.532 & 1.32 & 0.65 & 392 & 2.19 & 1.22 & 4.07 & 467 & 0.641 & 3.18\tablenotemark{b} & 0.38 & bright \\
C3 & 245.431 & -50.646 & 1.95 & 0.47 & 304 & 2.97 & 1.66 & 12.05 & 307 & $<$0.112 & $<$2.73\tablenotemark{d} & 0.86 & ... \\
C4 & 245.563 & -50.542 & 2.14 & 0.85 & 1088 & 2.38 & 1.33 & 6.65 & 348 & $<$0.116 & $<$2.74\tablenotemark{d} & 0.83 & ... \\
C5 & 245.314 & -50.665 & 3.97 & 0.96 & 2558 & 2.09 & 1.17 & 16.83 & 154 & 284.421 & 4.77\tablenotemark{b} & 3.46 & bright \\
C6 & 245.530 & -50.565 & 2.28 & 0.53 & 451 & 2.28 & 1.27 & 16.09 & 179 & $<$0.139 & $<$2.79\tablenotemark{d} & 1.72 & ... \\
C7 & 245.480 & -50.570 & 2.57 & 0.81 & 1173 & 2.71 & 1.51 & 13.20 & 224 & $<$0.337 & $<$3.02\tablenotemark{d} & 1.54 & ... \\
C8 & 245.515 & -50.547 & 2.28 & 0.48 & 361 & 2.48 & 1.39 & 9.25 & 359 & $<$0.544 & $<$3.14\tablenotemark{d} & 0.85 & ... \\
C9 & 245.162 & -50.738 & 4.57 & 0.88 & 2467 & 2.81 & 1.57 & 21.88 & 179 & 17.716 & 4.05\tablenotemark{b} & 3.43 & bright \\
C10 & 245.479 & -50.518 & 2.43 & 1.26 & 2682 & 3.10 & 1.73 & 7.78 & 339 & $<$0.083 & $<$2.65\tablenotemark{d} & 0.96 & ... \\
C11 & 245.398 & -50.567 & 1.51 & 0.55 & 321 & 3.86 & 2.16 & 8.91 & 441 & $<$0.275 & $<$2.96\tablenotemark{d} & 0.46 & ... \\
C12 & 245.539 & -50.460 & 1.66 & 0.72 & 605 & 3.24 & 1.81 & 6.28 & 480 & $<$0.320 & $<$3.00\tablenotemark{d} & 0.47 & ... \\
C13 & 245.403 & -50.520 & 2.23 & 1.08 & 1827 & 3.94 & 2.20 & 15.62 & 222 & $<$1.439 & $<$3.39\tablenotemark{d} & 1.36 & ... \\
C14 & 245.303 & -50.569 & 5.61 & 0.55 & 1178 & 3.43 & 1.91 & 10.83 & 621 & $<$0.917 & $<$3.28\tablenotemark{d} & 1.22 & ... \\
C15 & 245.114 & -50.685 & 3.57 & 0.60 & 897 & 2.90 & 1.62 & 24.79 & 174 & 38.016 & 4.25\tablenotemark{b} & 2.76 & bright \\
C16 & 245.606 & -50.387 & 1.64 & 0.49 & 280 & 3.57 & 1.99 & 9.94 & 400 & 0.050 & 2.52\tablenotemark{b} & 0.55 & faint \\
C17 & 245.323 & -50.508 & 3.55 & 1.07 & 2826 & 2.76 & 1.54 & 6.78 & 454 & 8.936 & 3.87\tablenotemark{b} & 1.05 & bright \\
C18 & 245.653 & -50.311 & 1.68 & 0.60 & 421 & 3.96 & 2.21 & 11.71 & 346 & $<$0.220 & $<$2.91\tablenotemark{d} & 0.65 & ... \\
C19 & 245.274 & -50.526 & 2.99 & 0.83 & 1447 & 3.75 & 2.09 & 16.05 & 271 & $<$0.379 & $<$3.05\tablenotemark{d} & 1.49 & ... \\
C20 & 245.591 & -50.309 & 1.95 & 1.06 & 1521 & 2.67 & 1.49 & 3.40 & 654 & 0.387 & 3.05\tablenotemark{b} & 0.40 & bright \\
C21 & 245.523 & -50.349 & 2.52 & 0.78 & 1063 & 3.82 & 2.14 & 6.44 & 655 & 0.069 & 2.60\tablenotemark{b} & 0.52 & faint \\
C22 & 245.261 & -50.488 & 2.53 & 0.73 & 952 & 3.14 & 1.75 & 15.03 & 238 & 3.113 & 3.60\tablenotemark{b} & 1.44 & bright \\
C23 & 245.125 & -50.561 & 2.43 & 0.87 & 1280 & 2.54 & 1.42 & 7.92 & 329 & 5.219 & 3.73\tablenotemark{b} & 1.00 & bright \\
C24 & 245.059 & -50.558 & 1.42 & 0.49 & 242 & 3.41 & 1.91 & 11.52 & 307 & $<$0.062 & $<$2.57\tablenotemark{d} & 0.62 & ... \\
C25 & 245.444 & -50.319 & 3.17 & 0.49 & 541 & 4.95 & 2.76 & 4.88 & 1573 & 0.229 & 2.92\tablenotemark{b} & 0.27 & bright \\
C26 & 245.260 & -50.395 & 1.59 & 0.89 & 887 & 3.45 & 1.93 & 9.27 & 304 & $<$3.432 & $<$3.62\tablenotemark{d} & 0.70 & ... \\
\hline
S1 & 245.487 & -50.899 & 1.82 & 0.86 & 946 & 2.61 & 1.46 & 10.49 & 222 & $<$0.086 & $<$2.66\tablenotemark{d} & 1.11 & ... \\
S2 & 245.427 & -50.917 & 1.57 & 0.66 & 478 & 3.00 & 1.67 & 9.79 & 289 & $<$0.054 & $<$2.54\tablenotemark{d} & 0.73 & ... \\
S3 & 245.203 & -51.000 & 1.12 & 1.29 & 1299 & 2.47 & 1.38 & 8.98 & 157 & 1.167 & 3.34\tablenotemark{b} & 0.96 & bright \\
S4 & 245.067 & -51.051 & 1.46 & 0.97 & 962 & 2.00 & 1.12 & 8.68 & 173 & 3.212 & 3.60\tablenotemark{b} & 1.14 & bright \\
S5 & 245.401 & -50.879 & 1.15 & 0.61 & 299 & 2.84 & 1.58 & 8.17 & 291 & 1.736 & 3.44\tablenotemark{b} & 0.53 & bright \\
S6 & 245.325 & -50.880 & 2.53 & 1.86 & 6123 & 3.38 & 1.89 & 11.47 & 210 & 18.896 & 4.07\tablenotemark{b} & 1.62 & bright \\
S7 & 244.927 & -51.126 & 1.11 & 0.92 & 649 & 2.36 & 1.32 & 9.88 & 161 & $<$0.073 & $<$2.62\tablenotemark{d} & 0.93 & ... \\
S8 & 245.013 & -51.063 & 2.13 & 0.61 & 552 & 2.08 & 1.16 & 8.34 & 285 & 5.675 & 3.75\tablenotemark{b} & 1.01 & bright \\
S9 & 244.981 & -51.065 & 2.56 & 0.62 & 687 & 1.90 & 1.06 & 6.80 & 347 & 13.061 & 3.97\tablenotemark{b} & 0.99 & bright \\
S10 & 245.079 & -51.004 & 1.19 & 0.52 & 226 & 2.06 & 1.15 & 10.79 & 176 & 0.422 & 3.08\tablenotemark{b} & 0.91 & faint \\
S11 & 244.792 & -51.155 & 1.48 & 0.66 & 452 & 1.81 & 1.01 & 9.55 & 174 & $<$0.053 & $<$2.54\tablenotemark{d} & 1.15 & ... \\
S12 & 245.340 & -50.821 & 1.55 & 0.94 & 957 & 3.00 & 1.68 & 16.30 & 145 & 1.094 & 3.32\tablenotemark{b} & 1.44 & bright \\
S13 & 244.902 & -51.057 & 5.04 & 1.01 & 3568 & 2.38 & 1.33 & 16.21 & 201 & 657.000 & 4.99\tablenotemark{c} & 3.38 & bright \\
S14 & 245.024 & -51.000 & 2.68 & 0.73 & 1001 & 1.51 & 0.85 & 11.28 & 157 & 1.760 & 3.45\tablenotemark{b} & 2.30 & bright \\
S15 & 244.790 & -51.107 & 2.81 & 0.51 & 508 & 1.73 & 0.97 & 14.42 & 173 & 4.466 & 3.69\tablenotemark{b} & 2.19 & bright \\
S16 & 244.844 & -51.072 & 1.61 & 0.50 & 282 & 1.75 & 0.98 & 15.91 & 120 & $<$0.018 & $<$2.26\tablenotemark{d} & 1.80 & ... \\
S17 & 245.065 & -50.943 & 1.51 & 0.49 & 251 & 3.96 & 2.21 & 17.84 & 240 & $<$3.759 & $<$3.65\tablenotemark{d} & 0.85 & ... \\
S18 & 245.022 & -50.954 & 2.50 & 0.72 & 897 & 2.62 & 1.46 & 15.68 & 191 & $<$1.874 & $<$3.46\tablenotemark{d} & 1.76 & ... \\
S19 & 244.783 & -51.069 & 2.86 & 0.65 & 836 & 1.32 & 0.74 & 14.55 & 117 & 30.058 & 4.19\tablenotemark{b} & 3.30 & bright \\
S20 & 245.147 & -50.859 & 1.97 & 0.69 & 650 & 2.12 & 1.18 & 9.14 & 241 & 2.924 & 3.58\tablenotemark{b} & 1.11 & bright \\
S21 & 245.035 & -50.891 & 2.87 & 2.29 & 10557 & 2.80 & 1.57 & 12.50 & 153 & 580.714 & 4.96\tablenotemark{c} & 2.52 & bright \\
S22 & 245.137 & -50.834 & 2.05 & 0.59 & 506 & 2.38 & 1.33 & 6.14 & 442 & 0.420 & 3.07\tablenotemark{b} & 0.63 & bright \\
S23 & 245.039 & -50.808 & 1.24 & 0.64 & 351 & 3.55 & 1.98 & 13.93 & 218 & $<$0.947 & $<$3.29\tablenotemark{d} & 0.77 & ... \\
S24 & 244.866 & -50.871 & 1.30 & 1.40 & 1781 & 1.56 & 0.87 & 8.60 & 106 & 0.560 & 3.15\tablenotemark{b} & 1.64 & bright \\
\enddata
\tablenotetext{a}{Estimated from the average ratio of the $\sigma_{v}(^{13}{\rm CO})/\sigma_{v}(\rm C^{18}O)$.}
\tablenotetext{b}{Estimated from the 12 and 22 $\mu$m fluxes (AllWISE) and the PACS 70 and 160 $\mu$m fluxes \citep{Chen_1995}.}
\tablenotetext{c}{Estimated from the IRAS fluxes 12, 25, and 60 $\mu$m \citep{Carpenter_2000}. }
\tablenotetext{d}{Estimated from the linear relation; ${\rm log}_{\rm 10}(L_{\rm bol}/L_{\rm \sun})=(0.60\pm0.08)\times {\rm log}_{\rm 10}\left(F_{\rm mid-IR}/{\rm Jy}\right) + (3.3 \pm 0.1)$. See Section 3.1.2.}
\end{deluxetable}

\end{document}